\documentclass[final,5p,fleqn]{elsarticle}
\usepackage{epsfig}
\usepackage{subfigure}
\usepackage[utf8]{inputenc}
\usepackage{float}
\usepackage{graphicx}
\usepackage{amsmath}
\usepackage{verbatim}
\usepackage{xcolor}
\usepackage{caption}
\usepackage{color}
\usepackage{soul}
\usepackage{amssymb}
\def\simge{\mathrel{%
       \rlap{\raise 0.511ex \hbox{$>$}}{\lower 0.511ex \hbox{$\sim$}}}}
\def\simle{\mathrel{
       \rlap{\raise 0.511ex \hbox{$<$}}{\lower 0.511ex \hbox{$\sim$}}}}

\newcommand \beq{\begin{eqnarray}}
\newcommand \eeq{\end{eqnarray}}

\usepackage[colorlinks=true,linktocpage=true,linkcolor=blue,citecolor=blue]{hyperref}

\begin{document}
\title{Macroscopic neutrinoless double beta decay: long range quantum coherence}
\author{Gordon Baym and Jen-Chieh Peng}
\address{\mbox{Illinois Center for Advanced Studies of the
Universe}  \\ \mbox{and Department of Physics, University of Illinois, 1110
  W. Green Street, Urbana, IL 61801} \\
}

\begin{abstract}
We re-introduce, in light of our modern understanding of neutrinos, the concept of ``macroscopic neutrinoless double beta decay" (MDBD) for Majorana neutrinos.  In this process
an antineutrino produced by a nucleus undergoing beta decay, $X\to Y + e^- + \bar \nu_e$, is absorbed as a neutrino
by another identical $X$ nucleus via the inverse beta decay reaction,   $\nu_e + X \to e^-+Y$.    The distinct signature of MDBD is that the total kinetic energy of the two electrons equals twice the endpoint energy of single beta decay.   The amplitude for MDBD, a coherent sum over the contribution of different mass states of the intermediate neutrinos, reflects quantum coherence over macroscopic distances, and is a new macroscopic quantum effect.   We evaluate the rate of MDBD for a macroscopic sample of ``$X$" material, e.g., tritium, acting both as the source and the target.
The accidental background for MDBD originating from two separate single beta decays, which contains two final state neutrinos, can be readily rejected by measuring the energy of the detected two electrons.   We discuss the similarities and differences between the MDBD and conventional neutrinoless double beta decay.
 While MDBD is clearly not a viable replacement for traditional $0\nu$DBD 
experiments, analysis of the concept of MDBD offers new perspectives on the physics of neutrinoless double beta decays.
\end{abstract}

\maketitle

\section{Introduction}

  The neutrino, if a Majorana rather than a Dirac fermion, would be its own 
antiparticle.  A key experimental signature distinguishing Majorana neutrinos from Dirac 
neutrinos is nuclear neutrinoless double beta 
decay (0$\nu$DBD) \cite{Racah,Furry,Primakoff-Rosen},
\beq
(A,Z)   \to (A,Z+2) +  e^- + e^-.
\label{MDBD4a}
\eeq
Despite the importance of determining whether neutrinos are Majorana or Dirac, and despite a major experimental effort, only upper limits for 0$\nu$DBD have so far been obtained \cite{Dolinski,Elliott,Agostini}.   Ongoing and future experiments with multi-ton detectors will further improve the sensitivity in these searches \cite{Barabash}.

   The existence of a Majorana neutrino implies lepton-number non-conservation, hence physics beyond the Standard Model.  Majorana neutrinos have only two spin states:  when massless the left-handed state is a neutrino and the right-handed 
an antineutrino.    The ``sea-saw" mechanism for Majorana neutrinos can then provide a natural explanation for the very light neutrino
masses inferred from tritium beta decay (TBD) and neutrino oscillation experiments \cite{Bala}.

 \begin{figure}[t]
\includegraphics*[width=1\linewidth]{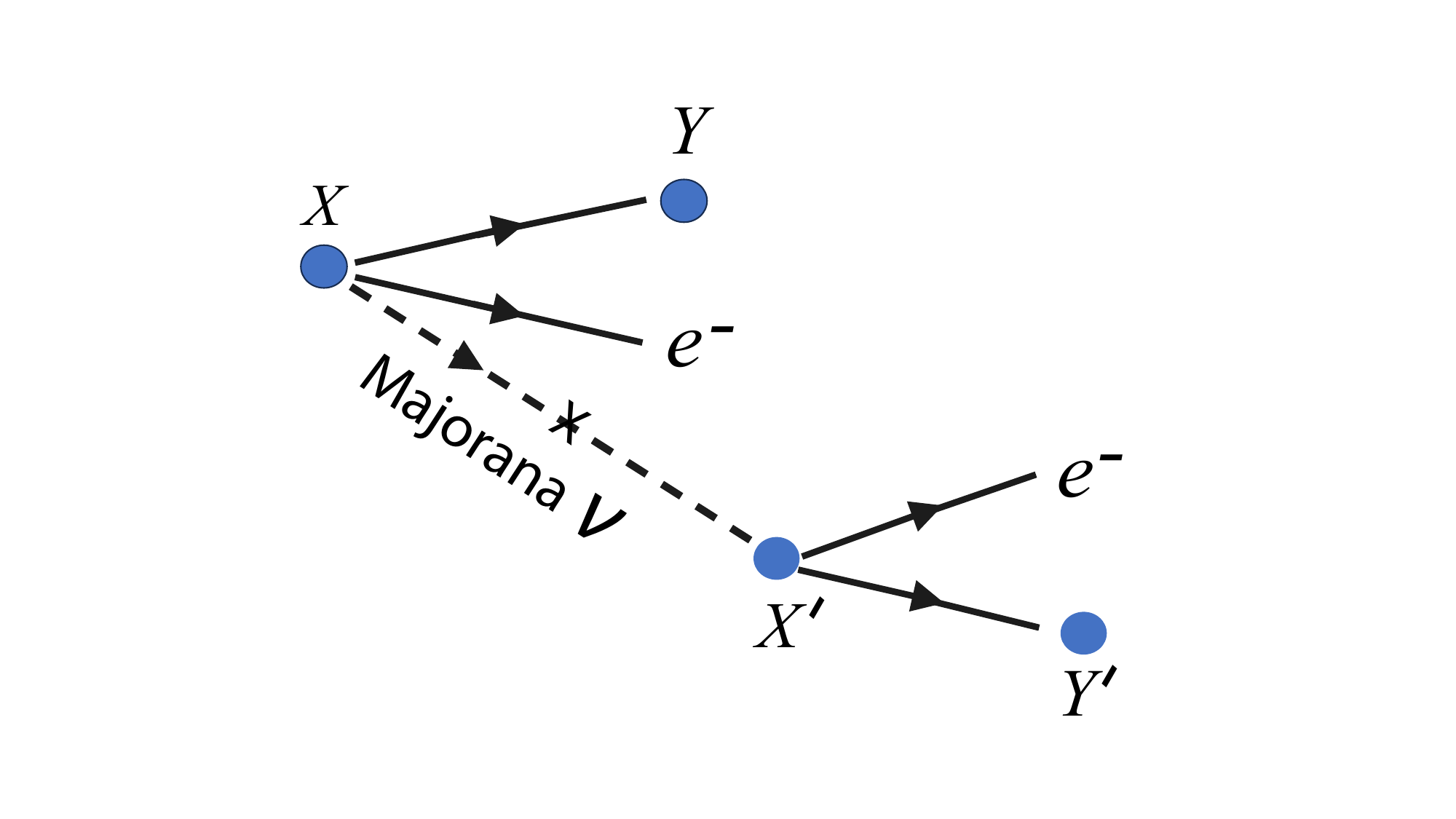}
\caption{
Illustration of macroscopic neutrinoless double beta decay, with time running to the right.  The first nucleus, $X$ undergoes a beta decay emitting an electron and a Majorana antineutrino, which is absorbed as a neutrino by a second nucleus $X'$ in an inverse beta decay.  The sum of the energies of the two electron emitted is just the sum of the individual endpoint energies in the beta decays of $X$ and $X'$.
}
\label{mdbdfig}
\end{figure}

  The standard picture of neutrinoless double beta decay is that the 
{\em antineutrino} emitted from a neutron in a nucleus is absorbed 
as a {\em neutrino} on a neutron in the same nucleus.  But there is in 
general no requirement in neutrinoless double beta decay that the 
second neutron be in the same nucleus -- as was recognized early 
in Ref. \cite{Pacheco} (and followed up in 
Refs. \cite{Kohyama, Skalsey}) -- or indeed that the second nucleus 
be the same nuclide as the first, or even that the two down quarks 
undergoing weak interactions be in different neutrons. 

  In this paper, we analyze in detail the concept of ``macroscopic neutrinoless 
double beta decay" (MDBD), in which a Majorana neutrino emitted from 
one nucleus is absorbed in a second nucleus.  In the single beta decay (XBD) from parent nucleus $X = (A,Z) $ to daughter nucleus $Y = (A,Z+1)$,
\beq
X\to Y + e^- + \bar \nu_e,
\label{MDBD1}
\eeq
the antineutrino produced is, for Dirac type neutrinos, distinct from $\nu_e$ and cannot
participate in the inverse beta decay (IXBD) reaction
\beq
\nu_e + X \to e^- + Y;
\label{MDBD2}
\eeq
the nucleus $X$ can only absorb an electron neutrino, $\nu_e$, to reach the final state $e^-+Y$.  However,
if neutrinos are Majorana, the neutrino and antineutrino are not distinct particles, and
the antineutrino in the XBD can then participate
in the IXBD.    The combination of these two sequential reactions, (\ref{MDBD1}) and
(\ref{MDBD2}), is the macroscopic neutrinoless double beta decay, illustrated in Fig.~\ref{mdbdfig},
\beq
X+ X' \to Y+Y' + e^- + e^-,
\label{MDBD4}
\eeq
where we add a prime to indicate the freedom of the two nuclei being different nuclides.  This process is
the macroscopic analog of the conventional neutrinoless double beta 
decay, $0\nu$DBD.
Both decays effectively turn two neutrons into two protons plus two
electrons. Both decays violate lepton number conservation and require 
the neutrino to be Majorana.  The signature of MDBD is a single
peak for the total energy of the two electrons located
at twice the endpoint energy of the single beta decay, analogous to the
single peak of the $0\nu$DBD decay just beyond the endpoint of the
$2\nu$DBD decay.
Detection of neutrinoless double beta decay, whether in a single nucleus or via MDBD exchange of a non-zero mass Majorana neutrino between different nuclei would violate lepton number conservation and would establish that neutrinos are Majorana and not Dirac.  

   Our intention is not to suggest realistic experiments -- they are 
not feasible within today's technology -- but rather to explore and 
stimulate new ways of looking at neutrinoless weak interactions.     
By analyzing neutrinoless weak interactions more broadly than are 
explored in  0$\nu$DBD  experiments, we gain new perspectives on the physics possibilities.
 
   For example, since in
neutrinoless weak interactions the initial source and final target 
exchange a Majorana neutrino in
various mass states, such exchange must exhibit {\em quantum interference} 
between different neutrino mass states.  This interference is implicitly 
included in calculation of microscopic  0$\nu$DBD \cite{Dolinski}.  
Macroscopic beta decay, however, opens new quantum interference effects 
over the macroscopic distance scales of the separation between 
source and target nuclei.

      MDBD and 0$\nu$DBD processes have notable differences.  
While only a limited number of nuclei are candidates for  0$\nu$DBD,
all $\beta$-unstable nuclei are potential candidates for MDBD. 
(The radioactivity of $\beta$-unstable nuclei, however, presents severe 
challenges to mounting a realistic experiment.)
 The uncertainties in the rate of MDBD are considerably smaller than 
in 0$\nu$DBD.  In 0$\nu$DBD  within a single nucleus, the uncertainty in 
the matrix elements \cite{Haxton,Engel} owes both to the many processes 
within the standard model that can contribute to the 
decay, e.g., $n\to \Delta^{++} + 2e$, where the $\Delta^{++}$ lives 
virtually, as well as exchanges of virtual particles beyond the 
standard model in addition to Majorana neutrino exchange.  
Beyond-the-standard-model physics that could lead to 0$\nu$DBD  
include right-handed weak currents, exchange of heavy neutrinos or 
supersymmetric particles, etc.\footnote{Nonetheless, even if such processes 
were to underlie  $0\nu$DBD, the observation of such double beta decays 
would establish that the neutrino is Majorana \cite{SV,Duerr}.} 
By contrast,  MDBD must come only from light massive Majorana neutrinos.  
Moreover, the matrix elements for sequential beta and inverse beta decay 
reactions in MDBD are precisely known.  The antineutrino emitted, 
as we detail below, does not need even to be an electron type, which 
raises the prospects of further neutrinoless weak interactions.  
Thus, for example, in $\pi^- \to \mu^- + \bar\nu_\mu$,
the muon antineutrino, if Majorana, can be absorbed as an electron 
neutrino in an inverse beta decay. 
     
  From an experimental perspective, while the 0$\nu$DBD rate is 
insensitive to the geometry of the source and is linearly proportional 
to the source mass, the rate for MDBD rate, as we spell out below, 
depends sensitively on the geometry of the source and is proportional 
to the source mass to the 4/3 power.  Moreover, one could in principle, as we 
discuss later, more readily
distinguish the accidental single beta decay background from the signal 
in MDBD processes than separating the 2$\nu$DBD background 
from the 0$\nu$DBD signal.
   
  In the next Section, \ref{rates}, we first review the calculation of the rate of single beta decay, and the cross section for inverse beta decay. In Sec.~\ref{sec-mdbd}  we derive the rate of MDBD as a coherent quantum process between the initial single beta decay and the inverse beta decay,  and derive the dependence of the MDBD rate on the geometry of the source.  Then in Sec.~\ref{background} we discuss how a MDBD signal is readily differentiated from the experimental two electron backgrounds.   We conclude in Sec.~\ref{conclusion}.

\section{Beta and inverse beta decay with massive neutrinos}
\label{rates}

  In this section we first review the role of chirality and helicity for massive neutrinos in beta and inverse beta decay.  We then briefly review the rates for emission of an antineutrino of given helicity in beta decay and the absorption of a neutrino of given helicity in inverse beta decay.
  
  \subsection{Chirality and helicity}
  
   Eigenstates of  $\gamma_5$ have definite chirality; we denote the $\gamma_5$ eigenvalue in these states  
  as {\em positive} ($\gamma_5 =1$) or {\em negative} ($\gamma_5 = -1$).   While chirality and helicity (the eigenvalue $h$ of the projection of the spin direction along the direction of motion, $\hat\Sigma\cdot \hat p$) coincide for massless neutrinos, where a positive chirality state always has positive helicity, $h$=1, and a negative chirality state always has negative helicity  $h$=\,-1,  they are not the same for finite mass neutrinos.  We denote positive helicity as  {\em right handed}, and negative helicity as {\em left handed}.  

   In beta decay, antineutrinos are emitted with positive chirality, $\gamma_5=1$, while neutrinos can be absorbed in the inverse reaction only if they have negative chirality.   Neutrinos are preferentially absorbed with helicity opposite to that of the antineutrino in beta decay.
 Since chirality is conserved for massless neutrinos, 
a massless Majorana antineutrino emitted in beta decay has the wrong chirality ever to be absorbed as a neutrino;  both MDBD and $0\nu$DBD are forbidden.  

    Independent of the neutrino mass, however, neutrinos and antineutrinos, in the absence of external forces, e.g., gravitational \cite{nugrav} or magnetic \cite{numag}, propagate in states of definite helicity.      When neutrinos have finite mass, chirality is no longer a constant of the motion, since 
$\gamma_5$ does not commute with the mass term in the Hamiltonian.  
In detail, as reviewed in the Appendix, an antineutrino emitted with positive chirality has amplitude $
  \alpha^\pm = \sqrt{(1\pm \beta_\nu)/2}
$
 to be right or left handed, where $\beta_\nu = v_\nu/c$, with $v_\nu$ the neutrino velocity.   Similarly, a right or left handed neutrino has amplitude $\sqrt{(1\mp \beta_\nu)/2}= \alpha^\mp$ 
to have the negative chirality needed to be absorbed in an inverse beta decay.   Thus a Majorana neutrino effectively has a total amplitude for emission with positive chirality and absorption with negative chirality, summed over both intermediate helicities, $\alpha^+\alpha^- + \alpha^-\alpha^+ =  \sqrt{1-\beta_\nu^2}= m_\nu/E_\nu$, where $m_\nu$ and $E_\nu$ are the neutrino mass and energy.   States of definite helicity have amplitudes to be of either chirality, thus allowing neutrinoless double beta decay for finite mass Majorana neutrinos.

\subsection{Single beta decay}

  We now discuss the amplitude, ${\cal A}^{ih}_{BD}$, and then the rate of beta decay, \eqref{MDBD1}, with emission of a $\bar\nu_e$ in mass state $i$ and helicity $h$, neglecting nuclear recoil (a better and better approximation the heavier the parent nucleus).  The amplitude  is 
\beq
   {\cal A}^{ih}_{BD} ={\cal M}_{i} \alpha^{ih}_{\bar\nu}.
   \label{bdampl}
\eeq
where $
  {\cal M}_{i} \propto  G_F V_{ud}  U_{ei}$,
with $G_F$ the Fermi weak coupling constant, $V_{ud}$ = 0.97425 the Cabbibo-Kobayashi-Maskawa (CKM) up-down quark matrix element, and  $U_{ei}$ the Pontecorvo-Maki-Nakagawa-Sakata (PMNS) neutrino mixing matrix element.   For right and left antineutrino helicities, $\alpha^{i,h=\pm 1}_{\bar\nu} = \alpha^\pm$.    For given neutrino energy, $\beta_i$ depends on the neutrino mass, $m_i$; since for $m_i\ll E_{\nu i}$, $\beta_i \simeq 1-m_i^2/2E_{\nu i}^2$, 
\beq
   \alpha^{iR}_{\bar\nu}\simeq 1, \,  \quad  \alpha^{iL}_{\bar\nu}\simeq \frac{m_i}{2E_{\nu i}}.
\eeq
 The dependence on the neutrino mass state is through the factors $U_{ei}$ and   $ \alpha^{ih}_{\bar\nu}$, and the helicity dependence resides in  $\alpha ^{ih}_{\bar\nu}$.

 The full squared matrix element has the form
\beq
   |{\cal M}_{i}|^2 = 2 \pi W(E_e)  |U_{ei}|^2, 
\eeq
where for tritium for example \cite{long,Hamish},
\beq
  W(E_e) = \frac{G_F^2}{2\pi} |V_{ud}|^2 F(Z,E_e)
  (m_{_{^3 \rm He}}/m_{_{^3\rm H}})|M_{f}|^2,  
\eeq
with $F(Z,E_e)$ the Fermi Coulomb factor for the electron-$^3$He system, and $|M_{f}|^2$ the sum of the nuclear form factors for Fermi and Gamow-Teller transitions.\footnote{For tritium, $|M_{f}|^2 = 1 + 2.788(g_A/g_V)^2$, with $(g_A/g_V)^2 = 1.6116$, while for the neutron, $|M_{f}|^2 = 1 + 3(g_A/g_V)^2$. }     In further calculation we symbolically write the matrix element as ${\cal M}_{i} = \sqrt{2\pi W } U_{ei}$.

  The rate of beta decay with emission of a neutrino of given helicity is then
\beq
  \hspace{-12pt}  d\Gamma_{XBD}^{h}  &=&\nonumber\\ && \hspace{-36pt} \sum_i \frac{d^3p_\nu}{(2\pi)^3}\frac{d^3p_e}{(2\pi)^3} 2\pi |{\cal M}_{i}|^2  (\alpha^{ih}_{\bar\nu})^2  \delta(\Delta M - E_e - E_{\nu i}  ), \nonumber\\
\label{TBDrate}
\eeq
 where $\Delta M=m_X-m_Y$ is the total energy released in the beta decay.

  Integrating over all electron states, and writing $d^3p_\nu = 4\pi p_\nu E_\nu dE_\nu$, 
we have
\beq
  \frac{d\Gamma_{XBD}^{h}}{dE_{\nu}}  &=& \frac{\bar\sigma(E_e)}{2\pi^2}  \sum_i|U_{ei}|^2 (1+h\beta_i) E_{\nu}\sqrt{E_\nu^2-m_i^2}, \nonumber\\
 \label{dgammade}
  \eeq
where $E_e= \Delta M - E_\nu $ and
\beq
  \bar\sigma(E_e) \equiv p_e E_eW(E_e).
\eeq
The differential rate of beta decay, for   $m_i \ll E_\nu$, summed over $h$ and using $\sum_i|U_{ei}|^2=1$,  is
\beq
  \frac{d\Gamma_{XBD}}{dE_{\nu}}  &\simeq&  \frac{\bar\sigma(E_e)}{\pi^2} E_\nu^2.
 \label{betaright}
\eeq
For tritium, in the limit of a zero energy neutrino  \cite{long}, $\bar\sigma\simeq 3.834 \times10^{-45}$ cm$^2$. 
The inverse of the mean lifetime, $\tau$, of the nucleus $X$ against beta decay (17.8 yr for tritium) is then 
$\int dE_{\nu}(d\Gamma_{XBD}/dE_{\nu})$.

\subsection{Inverse beta decay}

  While the rate of beta decay is an incoherent sum over the mass states in the emitted antineutrino,  
the rate of inverse beta decay depends strongly on the nature of the neutrino impinging on the target nucleus.  For example, neutrinos from the Big Bang arrive at Earth in definite mass eigenstates, a result of the spatial separation of the mass components of a neutrino wavepacket caused by the dependence of the neutrino velocity on the mass state.   By contrast, neutrinos from nuclear weak interactions, in $^{51}$Cr for example \cite{cr51}, are emitted in pure electron neutrino flavor states.  These neutrinos can initiate inverse beta decay whether the neutrino is Dirac or Majorana.   On the other hand, antineutrinos can participate in inverse beta decay (with electron emission) only if the neutrino is Majorana.  These antineutrinos can be either in mass eigenstates, e.g., if they come from the Big Bang, or in flavor eigenstates, e.g., from tritium beta decay or a nuclear power plant.
If the incident neutrino was emitted as a Majorana antineutrino, the amplitude for inverse beta decay is a coherent sum over the mass eigenstates in the neutrino. 

  Common to all these processes is the inverse beta decay amplitude for a neutrino in a given mass state.     Specifically, the amplitude, 
 ${\cal A}^{ih}_{IXBD}$, for inverse beta decay induced by a neutrino with velocity $\beta_i$, mass $m_i$ and helicity $h$ incident on a target nucleus, is 
\beq
   {\cal A}^{ih}_{IXBD} =  {\cal M}_{i}  \alpha^{ih}_{\nu},
  \label{ibdampl}
\eeq
 where $\alpha^{ih}_{\nu}= \sqrt{(1 -h \beta_i)/2}$, with ${\cal M}_{i} $ the same  as before.  
 
  The cross section for inverse beta decay for an incident neutrino in a given mass state $i$ and helicity $h$ is similarly (cf.~Eq.~\eqref{TBDrate}),  
\beq
  \frac{d\sigma_{IXBD}^{ih}}{d\Omega_e} &=&\nonumber\\&& \hspace{-36pt}\int\frac{p_e^2 dp_e}{(2\pi)^2} |{\cal M}_{i}|^2  (\alpha^{ih}_\nu )^2 \delta(m_X+E_\nu  -m_Y - E_e)  \nonumber\\ 
   && \hspace{-40pt}  = \frac{p_e E_e }{(2\pi)^2} |{\cal M}_{i}|^2 (\alpha^{ih}_\nu)^2  = \frac{ \bar \sigma(E_e)}{4\pi}   |U_{ei}|^2  (1-h \beta_i), 
   \nonumber\\
\eeq
and the total IXBD cross section $\sigma_{IXBD}^{ih}$ equals $\bar \sigma  |U_{ei}|^2   (1-h \beta_i)$.   As in beta decay, the helicity dependence is in the factor 
$\alpha^{ih}_\nu$.
 The scale of both the lifetime of a nucleus under beta decay, and the IXBD cross section is set by  $\bar \sigma(E_e)$.    

   If the initial neutrino is fully relativistic in mass state $i$  ($\beta_\nu \to 1$), then only the left handed helicity component can lead to emission of an electron, and the cross section for an incident mass state $i$ is $2\bar \sigma  |U_{ei}|^2$.  (Were the initial neutrino in an incoherent sum over mass states $i$, the cross section would bcome $2\bar\sigma$.)  But
for slowly moving
relic neutrinos ($\beta_\nu \to 0$), $\alpha^{ih}_\nu \to 1/\sqrt2$ and the IXBD amplitude becomes
independent of the neutrino helicity.   More generally, if the initial neutrino state is a coherent sum of mass states, the rate of IXBD reflects interference between the states, a problem we focus on in calculating the rate of MDBD below.

\section{Macroscopic double beta decay}
\label{sec-mdbd}

  We turn now to calculating the rate of macroscopic double beta decay in terms of the known physics of the single beta decay, XBD, and inverse beta decay, IXBD.   The rate of MDBD within a sample of atoms of nucleus $X$, with the sample acting both as the source of beta decays as well as the target for inverse beta decays, depends on both the microscopics of XBD and IXBD as well as the geometry of  the ensemble of $X$ atoms present.    For spacing between the atoms large compared with the characteristic intermediate neutrino wavelength, the neutrinos can be taken to be on-shell, in contrast to neutrinoless double beta decay within a single nucleus.
The MDBD process can take place for either helicity of the intermediate neutrino.

  Since the Majorana neutrino in MDBD exists only as an intermediary between the initial electron and final electron emission processes,  the contributions of the different neutrino mass states are coherent, and give rise to quantum interference.   To see this effect it is necessary to evaluate the MDBD rate in terms of the total amplitude for the process.   
  
  While neutrino oscillations  become important in processes over astrophysical distances, e.g., in the sun and in neutron stars, they are insignificant in MDBD over laboratory distances, except for very small neutrino energies.     
The characteristic length (in meters) in neutrino oscillations for a neutrino of energy $E_\nu$ (in MeV) is $
   L({\rm m})_{osc} \simeq   E_\nu({\rm MeV})/\Delta m_{31}^2 ({\rm eV}^2)$,
where $\Delta m_{31}^2 = m_3^2-m_1^2  \simeq 2.5\times 10^{-3} {\rm eV^2} $ corresponds to the difference of the squares of the masses of the heaviest and lightest neutrinos.   Thus $L_{osc} \sim  E_\nu$(MeV)  km.  Neutrinos of sub-keV energy could undergo neutrino oscillations in MDBD in large samples, but their effect would be hard to observe.
    
     We assume the initial beta decay to take place at point $\vec r$ and the inverse beta decay at $\vec r\,'$.  The total amplitude for the process $X+ X \to Y+Y + e^- + e^{-'}$ is 
\beq
  {\cal A}_{MDBD}(\vec r-\vec r\,') &=& \nonumber\\&&\hspace{-90pt}\sum_{ih} \int\frac{d^3p_\nu}{(2\pi)^3}  \frac{\left({\cal A}^{ih}_{XBD}e^{i (\vec p_\nu+\vec p_e) \cdot \vec r} \right)\left(  {\cal A}^{ih}_{IXBD}e^{i(\vec p\,_{e}' - \vec p_\nu )\cdot \vec r\,'}\right)}{\Delta M -E_e-E_\nu+i\eta},\nonumber\\
\eeq   
where $\eta\to 0^+$.
Then using the amplitudes~\eqref{bdampl} and \eqref{ibdampl}, doing the sum over helicities, 
\beq
\sum_h  \alpha^{ih}_{\bar\nu}\alpha^{ih}_{\nu} = \sqrt{1-\beta_i^2} = m_{\nu i}/E_\nu, 
\eeq
and the angular integration over the directions of the intermediate neutrino, we have
\beq
  {\cal A}_{MDBD} &=&   e^{i(\vec p_e\cdot \vec r+ \vec p\,_{e}' \cdot \vec r\,' ) } 
    \frac{\sqrt{W(E_e)W(E_e')}}{\pi |\vec r- \vec r\,'|} \sum_{i}U_{ei}^2m_{\nu i} 
  \nonumber\\ &&\times \int
\frac{E_\nu dE_\nu  \sin( p_\nu |\vec r- \vec r\,'|)}{\Delta M -E_e-E_\nu+i\eta}.   
\label{a117}
\eeq

   For large $|\vec r-\vec r\,'|$ compared with characteristic $1/p_\nu$, only the on-shell contribution of the neutrino (from the imaginary part $-i\pi \delta(X)$ of the denominator term, $1/(X+i\eta)$) is significant.  In this macroscopic limit the integral in the lower line of Eq.~\eqref{a117} becomes $-i\pi  E_\nu  \sin( p_\nu |\vec r- \vec r\,'|)$, where now $E_\nu = \Delta M -E_e$.  We drop the overall phase factors, which do not enter 
the MDBD rate, and finally arrive at the MDBD amplitude,   
\beq
  {\cal A}_{MDBD} &=& \sqrt{W(E_e) W(E_e') } \frac{\sin( p_\nu |\vec r- \vec r\,'|)}{ |\vec r- \vec r\,'|} \sum_{i} U_{ei}^2 m_{\nu i}.\nonumber\\
\eeq

  The total rate of MDBD with one emitter and one absorber is then
\beq
   \Gamma_{MDBD} &=& \nonumber\\ && \hspace{-48pt}\int \frac{d^3p_e}{(2\pi)^3} \frac{d^3p_e'}{(2\pi)^3} | {\cal A}_{MDBD} |^2  2\pi \delta(2\Delta M - E_e-E_e'). \nonumber\\
\eeq
For $p_\nu |\vec r- \vec r\,'| \gg 1$ the $\sin^2$ factor, slightly cross-grained in $\vec r\,'$, becomes 1/2, and thus
\beq
       \Gamma_{MDBD} &=&\frac{{\cal K}}{4\pi |\vec r- \vec r\,'|^2},
\eeq
where the microscopic rate factor, after doing the $E_e'$ integral, becomes
\beq
   {\cal K} &=& \bar m^2   \int \frac{\bar\sigma(E_e)\bar\sigma(E_e')}{\pi^2} dE_e.
   \label{calK}
\eeq
Here $E_e' = 2\Delta M - E_e$, and 
\beq
   \bar m & \equiv  |\sum_{i} U_{ei}^2 m_{\nu i}| 
\eeq
is the average neutrino mass entering neutrinoless double beta decay. 
 
   We can understand the remaining structure of the MDBD rate by considering
simply the single mass state contribution, denoted as $\Gamma_{MDBD}^{\,i}$.
   The beta decay of an $X$ nucleus at point $\vec r$ produces a flux,  per unit neutrino energy, of 
neutrinos in helicity state ${h}$ of energy $E_\nu$ at point $\vec r\,'$, 
\beq
    \frac{dF^{ih}(\vec r\,')}{dE_{\nu i}}  
= \frac{1}{4\pi|\vec r - \vec r\,'|^2}  
\frac{d\Gamma_{XBD}^{ih}}{dE_{\nu i}} .
\eeq  
Thus the full rate for neutrinos emitted at point $\vec r$ in mass state $i$ to be absorbed at point $\vec r\,'$ is 
\beq
    \Gamma_{MDBD}^{\,i} & = &\sum_{h} \int \frac{dF^{ih}(\vec r\,')}{dE_{\nu i}}dE_{\nu i} \sigma_{IXBD}^{ih}
=\frac{{\cal K}_i}{4\pi|\vec r - \vec r\,'|^2}, \nonumber\\
 \label{GMDBD}
\eeq
where 
\beq
{\cal K}_i &=& \int dE_{\nu i}  \sum_{h}\frac{d \Gamma^{ih}_{XBD}}{dE_{\nu i}}  \sigma^{ih}_{IXBD} \nonumber\\ 
 & =& |U_{ei}|^4 m_{\nu i}^2 \int dE_{\nu i}\frac{\bar\sigma(E_{e})\bar\sigma(E_{e}')}{\pi^2};
 \eeq 
this result is precisely the single neutrino mass state ($i$) contribution to $\cal K$, Eq.~\eqref{calK}, with the identifications
$E_e = \Delta M - E_{\nu i}$ and $E_{e}' = \Delta M + E_{\nu i} = 2\Delta M -  E_{e}$.   

 The integral in $\cal K$ depends very weakly on the neutrino masses since $m_\nu \ll m_e$, and we neglect them in the integral.     Furthermore the $\bar\sigma$ do not vary greatly over the range of neutrino energies, and we write them as $\bar\sigma$ evaluated at the end point energy times a correction. 
Neglecting the energy dependence in the Fermi Coulomb term, $F$ in
 $\bar\sigma$ we have
 \beq
{\cal K}  & =& \frac1{\pi^2} \bar m^2 \bar \sigma_{end}^2 K_{end} {\cal J},   
\label{calKJ}
\eeq
where $K_{end}=\Delta M - m_e$ is the endpoint kinetic energy of the XBD, $\bar\sigma_{end} \equiv \bar\sigma(E_e=\Delta M)$,  and the dimensionless factor ${\cal J}$ is given by
\beq
  {\cal J}(m_e/\Delta M) &=& \int_0^{K_{end}} \frac{dE_{\nu}}{K_{end}} \frac{p_e E_e p_e' E_e'}{(K_{end}^2 - m_e^2)K_{end}^2}. \nonumber\\
  \eeq  
 
 \begin{figure}[t]
\vspace{-1.4in}
\includegraphics*[width=1.1\linewidth]{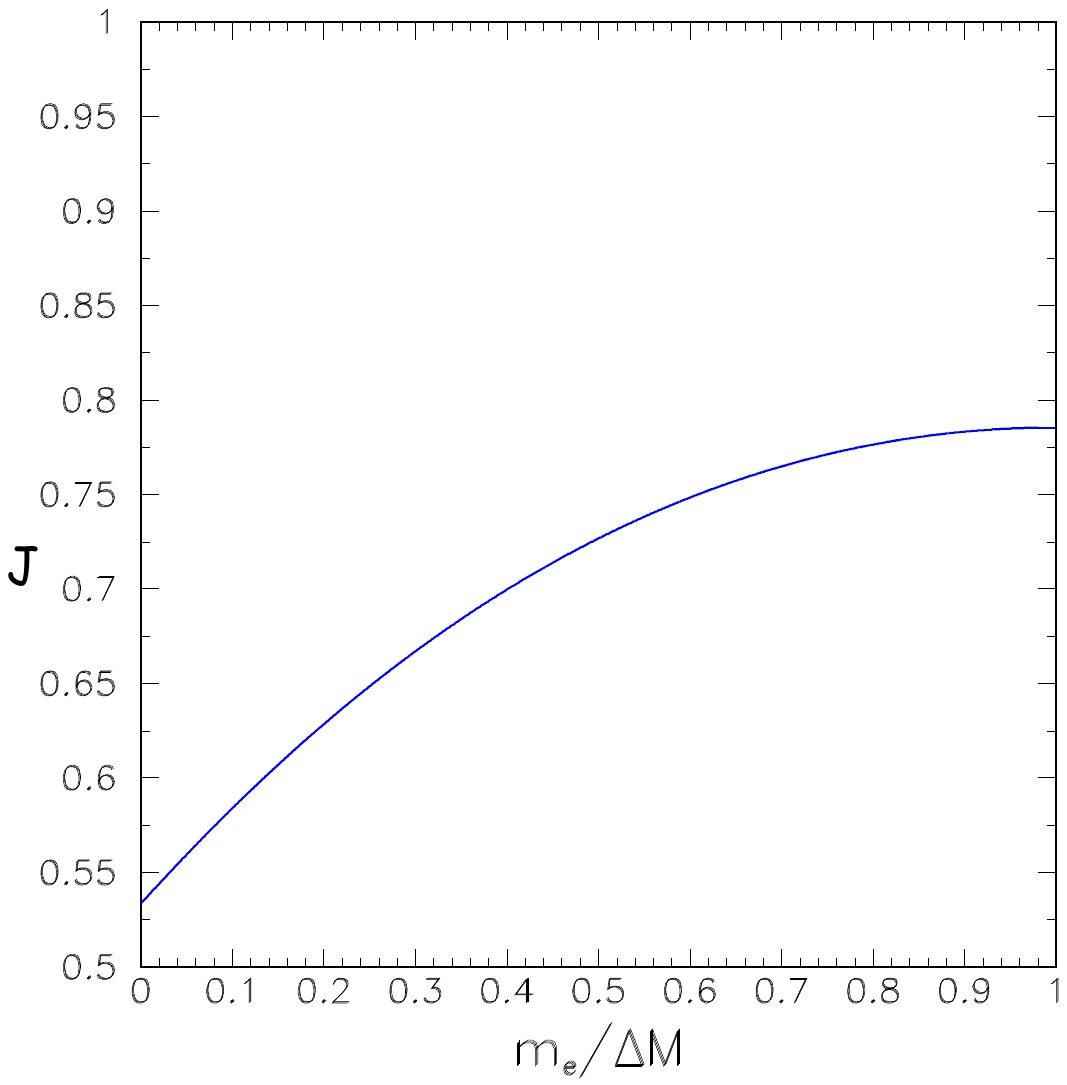}
\caption{The dimensionless integral $\cal J$ entering the MDBD rate, as a function of $m_e/\Delta M$.}
\label{calJ}
\end{figure}
 
  Figure~\ref{calJ} shows $\cal J$ as a function of $m_e/\Delta M$.   In the limit that the Q value in the XBD is small compared with the electron mass, or  $\Delta M -m_e << m_e$, as in tritium, then ${\cal J}\to \pi/4$ (this factor arises from the averaging of the momenta of the electrons), and
\beq
{\cal K} &\to & \frac{\bar m^2 \bar\sigma_{end}^2}{4\pi}  K_{end}.
\label{yisone}
 \eeq 
 In the opposite limit, $\Delta M \gg m_e$,  the correction factor ${\cal J}$ becomes 8/15, and ${\cal K} \to (8\bar m^2 \bar\sigma_{end}^2/15\pi^2) K_{end}$.

 \begin{figure}[t]
 \vspace{-1in}
\includegraphics*[width=1\linewidth]{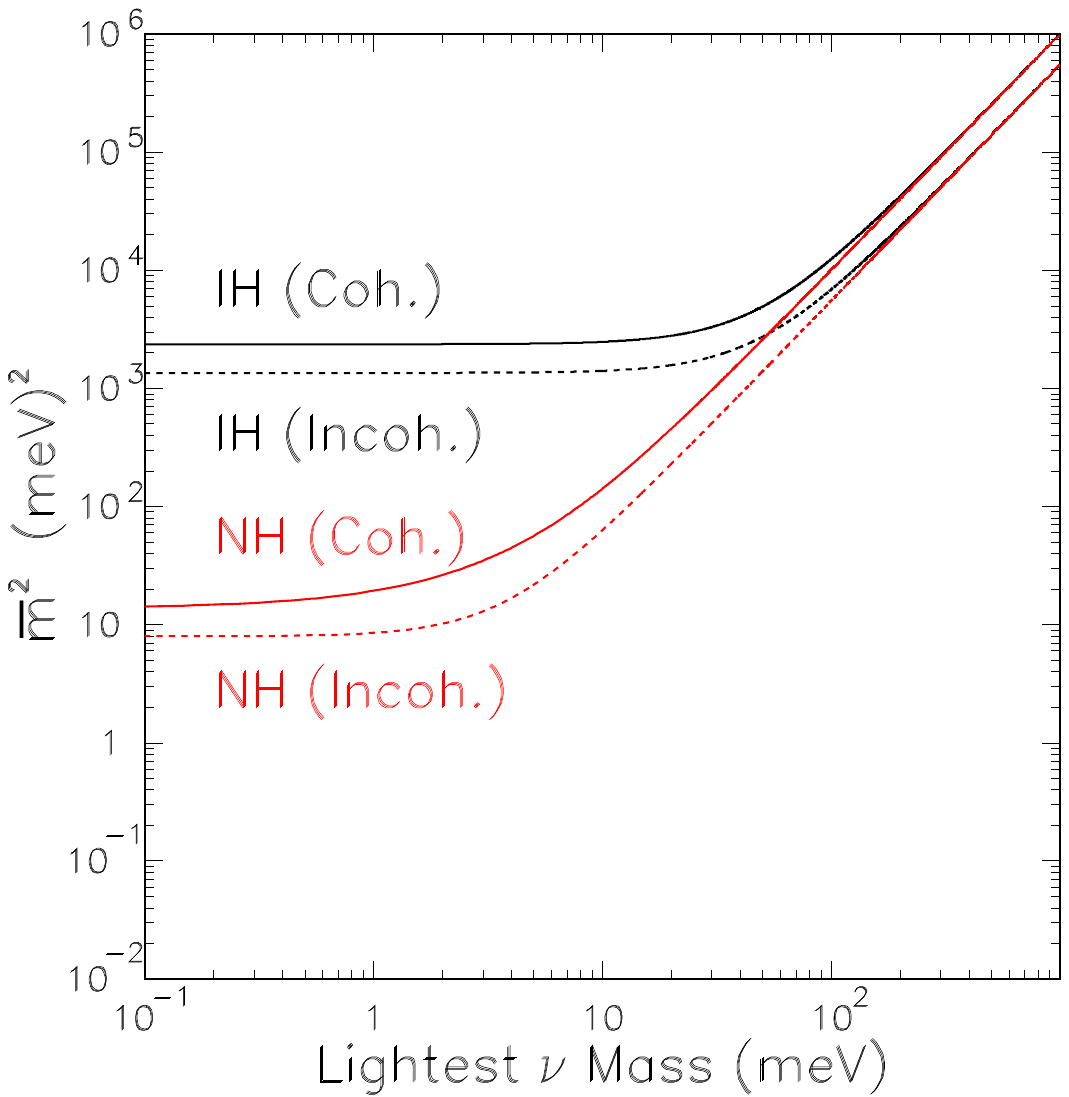}
\caption{
 The calculated $\bar m^2$ for the normal (NH) and inverted (IH) neutrino mass hierarchies with and without interference between the mass eigenstates.   The curves labelled ``coherent" assume zero Majorana phases (with respect to the CP phase), while the curves labelled ``incoherent"  neglect the off-diagonal interference between different mass eigenstates.  In this case coherence increases $\bar m^2$ and the rate of MDBD.}
\label{m2}
\end{figure}

\begin{figure}[t]
 \vspace{-0.60in}
\includegraphics*[width=1\linewidth]{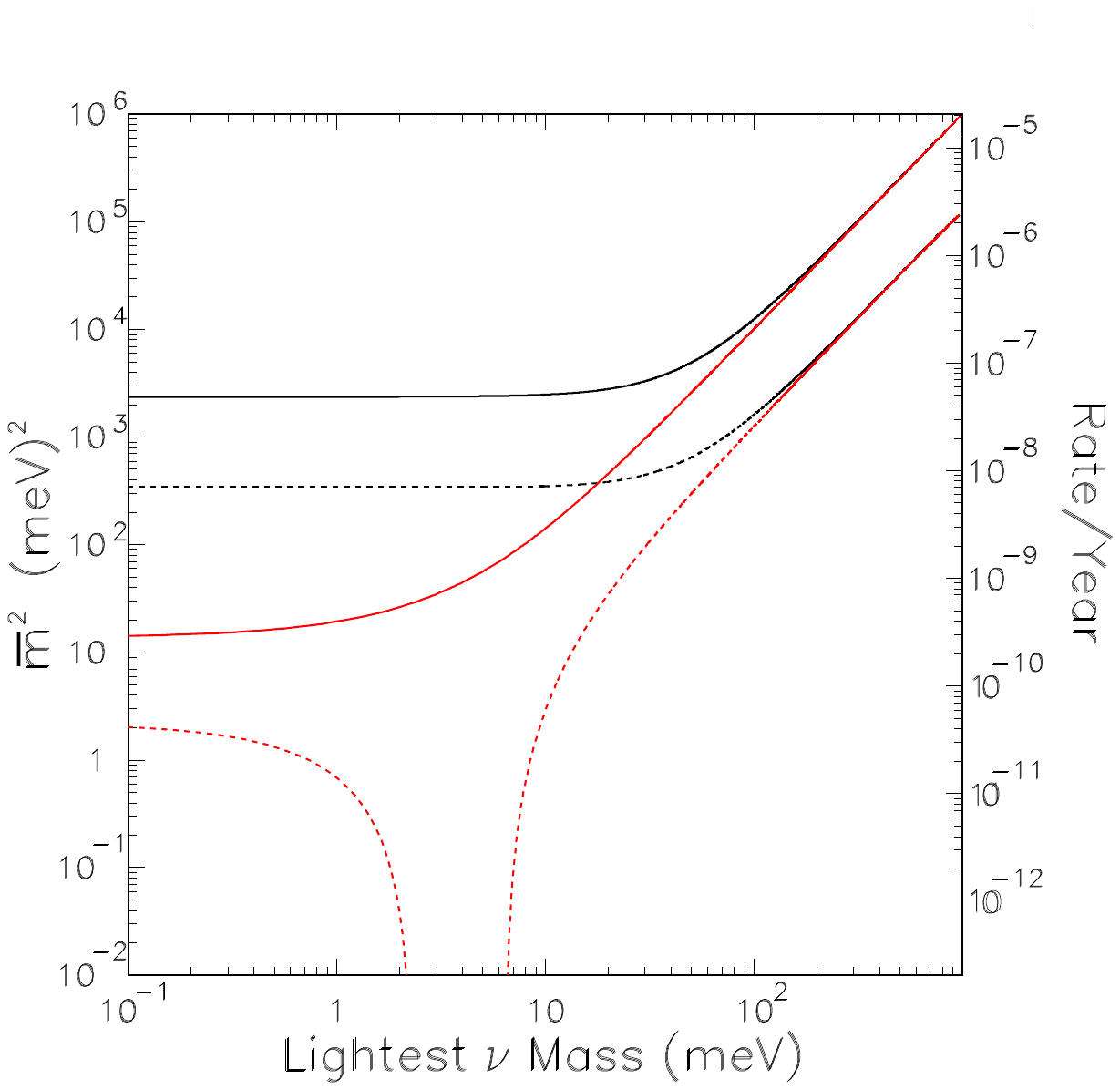}
\vspace{-41pt}
\caption{ Full range of $\bar m^2$ for the normal and inverted hierarchies.  The solid curves show the same curves labelled ``coherent" in Fig.~\ref{m2}, while the dashed curves show the maximal destructive interference from possible Majorana phases.   The scale on the right shows the corresponding MDBD rate expected in a 100 g tritium target.   The rate is proportional to the mass of the target to the 4/3 power. }
\label{m2interf}
\end{figure}

\subsection{Quantum interference and coherence}

 The rate of MDBD is  proportional to the square of the weighted neutrino mass  $|\sum_i  U_{ei}^2 m_{\nu i}|^2$.  This term includes 
both the direct ($i=j$) MDBD process, with a sum of the rates for single mass eigenstates, $\sum_i |U_{ei}|^4 m_{\nu i}^2$, as well as the interference between different mass eigenstates ($i\ne j$).   
   By contrast,  if the incident neutrino is an incoherent sum of mass states,   then the rate is proportional to the $i=j$ term alone, i.e., 
\beq
      \sum_i  |U_{ei}|^4m_{\nu i}^2 \equiv m_{incoh}^2,
 \eeq
a very different weighting of the neutrino masses.   

The dependence of $\bar m$ on possible phases of the $U_{ei}$ is a consequence of quantum interference between different mass states $i$
in macroscopic double beta decay.  Explicitly, $\bar m$ can be written in  terms of the CP phase $\delta$ and Majorana phases $\lambda_a$ and $\lambda_b$,  as,
 \beq
   \bar m =\big| |U_{e1}|^2 e^{2i\lambda_a}m_1 + |U_{e2}|^2 e^{2i\lambda_b} m_2+ |U_{e3}|^2 e^{-2i\delta}m_3\big|. \nonumber\\
 \label{MASS_NU} 
\eeq
 
    Figure~\ref{m2} shows $\bar m^2$  as a function of lightest neutrino mass for the normal and inverted mass hierarchies, using the neutrino oscillation parameters tabulated in~\cite{Qian}.
 The curves labelled ``incoherent" neglect the interference terms in $\bar m^2$, and include only the diagonal sum over mass eigenstates, i.e., with $\bar m^2$ replaced by $m_{incoh}^2$.   In the curves labelled ``coherent,"  we assume that the Majorana CP phases equal $-\delta$, so that all the phases drop out of $\bar m$, and 
$\bar m^2 - m_{incoh}^2 = \sum_{i\ne j}|U_{ei}|^2|U_{ej}|^2m_{\nu i}m_{\nu j} >0$. In this case, coherence increases the MDBD rate.    For given neutrino mass splittings, the curves are in fact simply quadratic functions of the lightest neutrino mass.  

 On the other hand, if the Majorana CP phases differ from $ -\delta$, coherence can, as illustrated in Fig.~\ref{m2interf}, decrease the rate, exactly as in 0$\nu$DBD~\cite{Dolinski}.     Figure~\ref{m2interf} also shows, on the right hand scale, the MDBD rate for a spherical source of 100 g of tritium, calculated from Eq.~\eqref{calR} below.

  Macroscopic double beta decay remains coherent as long as the neutrino wavepacket travelling from source to target stays together.  However, 
the different mass components of the wavepacket physically separate and destroy coherence at large distances owing to the mass dependence of their velocities, $\delta 
v/c\sim  \frac12 \Delta m_\nu^2/E_\nu^2$ \cite{numag}.   If the initial neutrino wavepacket is of length $\ell$, then decoherence takes place over distances $L_{dec} \sim \ell E_\nu^2/\Delta m_\nu^2 \sim  \ell E_\nu L_{osc}$, where $L_{osc}$, discussed above,  is the length over which neutrino oscillations become significant.    The estimate  $\ell E_\nu\sim 1$ nm-MeV \cite{jones} implies that $L_{dec}$ is $\sim 10^{4}L_{osc}$, or 
$L_{dec} \sim 10^4 E_\nu({\rm MeV})/\Delta m_\nu^2({\rm eV}^2)$ m.  A detailed discussion of decoherence over astrophysical distances must thus take flavor oscillations of the neutrino wave packets into account.
 
\subsection{Dependence of the MDBD rate on geometry}
\label{geometry}

  We next consider the rate of MDBD in a macroscopic sample of N atoms of $X$ with uniform density $n$.   For an ensemble of beta emitters, Eq.~\eqref{GMDBD} implies the total rate of MDBD,
\beq
 {\cal R}_{MDBD} &=& \int d^3r d^3r'\frac{n^2{\cal K}}{4\pi|\vec r - \vec r\,'|^2} = \frac{N^2 {\cal K}}{4\pi}  \Big\langle \frac1{(\vec r-\vec r\,')^2}\Big\rangle. \nonumber\\
 \label{GMDBD1}
\eeq  
For isotropic scaling at fixed density, the rate scales as the fourth 
power of the linear size of the system, or $N^{4/3}$, as first noted
by Pacheco~\cite{Pacheco}.
On the other hand, the rate of 0$\nu $DBD scales with $N$.  

   The detailed evaluation of the integral depends on the geometry of the target.  It is readily evaluated for a spherical target of radius $R$ by writing,
\beq
  \int \frac{d^3r  d^3 r'}{ (\vec r - \vec r\, ')^2 } =  \int\frac{d^3k}{4\pi k} \, \Big| \int d^3r e^{i\vec k\cdot \vec r}\,\Big|^2 = 4\pi^2 R^4,
\eeq
where we use  $\int d^3r e^{i\vec k\cdot \vec r} = (4\pi/k^3)(\sin kR - kR \cos kR)$.

  The total double beta decay rate for a spherical source\footnote{ We give the result for a spherical source only for illustration, ignoring the complication that electrons produced within a sufficiently large spherical sample rapidly lose energy traversing the source and cannot readily be detected.
For such a reason, the PTOLEMY experiment \cite{ptolemy} is aiming for two dimensional targets, using tritiated graphene.  A realistic MDBD experimental design would similarly require a more planar geometry, at the cost of reducing the probability of a given antineutrino from the initial beta decay finding a target for inverse decay.    } is then
\beq
       {\cal R}_{MDBD} =    \frac{9}{16\pi}  \frac{N^2}{R^2}   {\cal K},
   \label{calR}
\eeq
which shows explicitly the scaling with the size of the sample (at fixed density) as $N^{4/3}$.

\subsection{Estimate of  $R_{MDBD}$}

    We estimate $R_{MDBD}$ from Eq.~\eqref{calR} with \eqref{yisone}.  
For tritium,
\beq
       {\cal R}_{MDBD} = \frac{9}{64\pi^2}\frac{N_T^2}{R^2}  {\bar m}^2  
\bar\sigma_T^2 K_{end}\,;
\eeq
with a 100 g sphere of tritium ($N_T \simeq 2 \times 10^{25}$ nuclei), 
${\cal R}_{MDBD} \simeq 2.32 \times 10^{-5}(\bar m/1\, {\rm eV})^2$/year.
It is instructive to compare this rate
with the conventional neutrinoless double beta decay rates for various
nuclei. No $0\nu$DBD events have been positively identified \cite{Barabash},
and the most sensitive recent results include those on 
$^{76}$Ge~\cite{GERDA,Majoranaexpt}, $^{136}$Xe~\cite{kamLAND,xenonNT,EXO},
$^{130}$Te~\cite{CUORE}, $^{82}$Se~\cite{CUPID,NEMO}, 
and $^{100}$Mo~\cite{CUPID_Mo}. Table I lists the expected $0\nu$DBD rates 
for these nuclei with an exposure of 100 g-yr, assuming 
$\bar m$ = 0.1 eV.   As we see in the Table, the tritium MDBD rate is four to five orders of magnitude
lower than those expected for $0\nu$DBD.  This difference reflects the much smaller
Q=16.8 keV value in tritium beta decay than in
$0\nu$DBD, e.g., Q = 2.04 MeV for $^{76}$Ge and 
Q=2.46 MeV for $^{136}$Xe $0\nu$DBD. 

\begin{table*}
\begin{center}
\begin{tabular}{|c|c|c|}
\hline
\hline
Nucleus & $T_{1/2}$ for $\bar m =0.1$ eV & Yield per 100 g-yr\\
\hline
$^{3}$H (MDBD) & -- & $2.3 \times 10^{-7}$ \\
$n$ (MDBD) & -- & $3.4 \times 10^{-2}$ \\
$^{11}$C (MDBD) & -- & $5.1 \times 10^{-5}$ \\
$^{76}$Ge ($0\nu$DBD)~\cite{GERDA,Majoranaexpt} & $1.1 \times 10^{26} 
< T_{1/2} < 6.0 \times 10^{26} $ yr & 
$9.1 \times 10^{-4} < Y < 5.0\times 10^{-3}$\\
$^{136}$Xe ($0\nu$DBD)~\cite{kamLAND,xenonNT,EXO} & $3.0 \times 10^{25}
< T_{1/2} < 5.6 \times 10^{26} $ yr &
$5.5 \times 10^{-4} < Y < 1.0 \times 10^{-2}$  \\
$^{130}$Te ($0\nu$DBD)~\cite{CUORE} & $1.8 \times 10^{25}
< T_{1/2} < 2.0 \times 10^{26} $ yr &
$1.6 \times 10^{-3} < Y < 1.8\times 10^{-2}$ \\
$^{82}$Se ($0\nu$DBD)~\cite{CUPID,NEMO} & $3.2 \times 10^{25}
< T_{1/2} < 2.3 \times 10^{26} $ yr &
$2.3 \times 10^{-3} < Y < 1.6 \times 10^{-2}$\\
$^{100}$Mo ($0\nu$DBD)~\cite{CUPID_Mo} & $1.4\times 10^{25}
< T_{1/2} < 4.3 \times 10^{25} $ yr &
$9.7 \times 10^{-3} < Y <  3.0 \times 10^{-2}$ \\
\hline
\hline
\end{tabular}
\end{center}
\caption {
Expected yields, $Y$, for MDBD and $0\nu$DBD for a 100 g-yr
exposure using various nuclei.  We assume the Majorana neutrino effective mass $\bar m = 0.1$ eV.
The uncertainties in the predicted $T_{1/2}$ and yield for 0$\nu$DBD reflect the range of uncertainties of the nuclear matrix elements adopted by Ref.~\cite{Barabash} to relate the current experimental limits on $T_{1/2}$ to values of $\bar m$. 
 For MDBD we assume 100 g spherical
sources of tritium (and neutrons, for illustration) with density 1 g/cm$^3$ and 
$^{11}$C with density 2.2  g/cm$^3$.   }
\label{itbd_tab}
\end{table*}

   To illustrate the importance of a large Q value, we 
compare MDBD for neutrons (where $n \to p + e^- + \bar \nu_e$ followed 
by $\nu_e + n \to p + e^-$) with tritium.
In this double beta decay Q= 2 $\times 0.782$ MeV = 1.564 MeV.  
Furthermore $\bar \sigma_n$ is a factor $\sim 20$ larger than in tritium, 
which has comparable matrix elements.   In a sample 
of 100 g of neutrons occupying the spherical volume as would 100 g of 
tritium, the MDBD rate is a factor $\sim  
10^5$ larger than in 
a comparable sample of tritium. In fact, the MDBD rate for such a neutron
sample is comparable to that in an equivalent sample of nuclei in $0\nu$DBD.
Clearly though, it would be impossible 
to do such an experiment in a cloud of laboratory neutrons.\footnote{On the other hand, a neutron star contains beneath 
its crust a stable liquid of order $10^{57}$ neutrons in a $\sim$10 km 
sphere, which implies a geometric enhancement over MDBD in 100 g of 
tritium by a factor $\sim 10^{53}$.   The rate of MDBD is, however, 
significantly suppressed by Pauli exclusion in both URCA and modified 
URCA processes, as well as by BCS pairing.   Unfortunately there is 
no way apparent to distinguish such MDBD processes in neutron stars, 
even in their cooling.}

  We have also determined the MDBD rate for $^{11}$C, 
which undergoes a $\beta^+$ decay with a half life of 
22.3 min. and a relatively large Q value of 1.98 MeV. 
We estimate that a 100 g 
spherical sample of $^{11}$C, which has a density 2.2 g/cm$^3$, and 
contains 3/11 of the number of atoms in a 100 g sample of tritium, 
would have a MDBD rate of $5.14 \times 10^{-5}$ per year for 
$\bar m = 0.1$ eV, as shown in Table I.   This
rate is significantly larger than the MDBD rate for tritium, but still lower than the rate for the $0\nu$DBD sources 
listed in Table I.  As mentioned, the MDBD rate contains
no uncertainties arising from the nuclear matrix elements, distinctly different
from in $0\nu$DBD.  While considering the MDBD rate for $^{11}$C is instructive, we caution that the relatively short lifetime 
of $^{11}$C implies that the $^{11}$C would have decayed away
well prior to a significant MDBD signal.

\section{MDBD Signal vs. Background}
\label{background}

 \subsection{Electron distribution in MDBD}
 
   The characteristic feature of MDBD is that the sum of the two electron energies is just twice the total energy released in a single beta decay.   The distribution of the electron pairs has the structure, in the limit of small neutrino mass,
\beq
  &d^2N(E_e,E_e') &\propto p_e E_e F(E_e) dE_e p_e' E_e' F(E_e') dE_e'  \nonumber\\ &&\hspace{-36pt}\times \int  dE_\nu \delta(\Delta M -E_\nu - E_e)\delta(\Delta M +E_\nu - E_e') \nonumber\\ &&
   =  p_e E_e F(E_e)F(E_e') dE_e p_e' E_e' dE_e'  \nonumber\\ &&\hspace{48pt}\times\delta(2\Delta M -E_e - E_e'),
\label{mdbdspec}
\eeq 
where $E_e$ and  $p_e$ refer to the energy and momentum of
the electron emitted in the single beta decay process, and 
$E_e'$ and $p_e'$ to the electron emitted subsequently in the neutrino
capture process in MDBD. In Eq.~\eqref{mdbdspec}, the $E_\nu^2$ from 
the neutrino phase 
space is canceled by 
neutrino helicity factor $1-\beta^2 = (m_\nu/E_\nu)^2$.
Then the single electron distribution in MDBD has the form,
\beq
  \frac{dN_e}{dK_e} (MDBD)
  &\propto&  p_e E_e F(E_e)F(E_e')  p_e' E_e',
 \eeq
 \label{mdbdspec7}
where here $E_e' = 2\Delta M - E_e$.    

\begin{figure}[t]
\vspace{-1.6in}
\includegraphics*[width=1.1\linewidth]{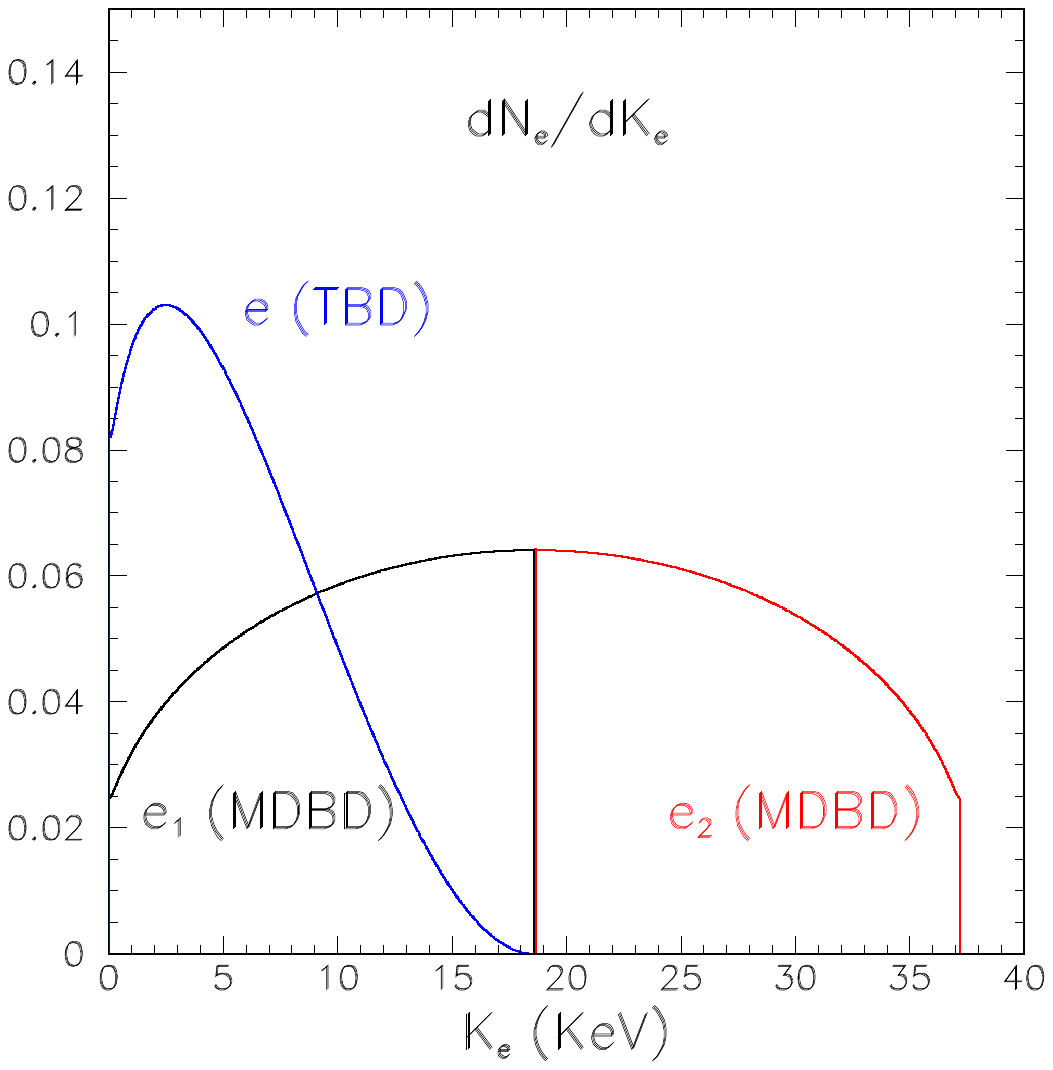}
\caption{The electron energy distributions in single tritium 
beta decay (TBD) and in tritium MDBD.  Here $e_1$ is 
the electron
produced in the initial beta decay, and $e_2$ the electron emitted
in the subsequent inverse beta decay. Electron $e_1$ has the lower energy, with
$K_e(max)$ equal to $K_{end}$, while $e_2$ has the higher energy
with $K_e(max)$ equal to twice of $K_{end}$.
The detailed structure of the 
single electron energy distributions at $K_{end}$, shown as a single 
vertical line, is in fact two curves separated essentially 
by $2m_\nu$, too fine to be seen in the figure.  
}
\label{edistr}
\end{figure}

The expected electron energy spectra are shown in Fig.~\ref{edistr} 
in terms of the electron kinetic energy $K_e$ for tritium MDBD.  Here
$e_1$ denotes the electron
produced in the initial beta decay, and $e_2$ the electron emitted
in the subsequent inverse beta decay. The energy spectra of $e_1$ and $e_2$
are clearly symmetric about the single beta decay endpoint, $K_e = K_{end}$,
according to Eq.~\eqref{mdbdspec}.   While the maximal energy for $e_1$ 
equals the end-point energy $K_{end}$, $e_2$ has a higher energy with 
$K_e(max)$ equal to twice of $K_{end}$.
Figure~\ref{edistr} also
shows for comparison the electron energy distribution from tritium single beta 
decay, which vanishes near $K_{end}$, as we see from Eq.~\eqref{betaright}, 
essentially as
\beq
   \frac{dN_e}{dK_e} (XBD) \propto \frac{d\Gamma}{dE_e} = \frac{\bar\sigma(E_e)}{2\pi^2} (\Delta M - m_e - K_e)^2;   
\eeq
the distribution is dominated by electrons emitted with a right 
handed neutrino.   In this figure, we normalize the individual 
distributions to unity, and take the Fermi Coulomb correction to be 
$2\pi\eta/(1-e^{-2\pi\eta})$ with $\eta = Ze^2/ v_e$.   
The non-zero neutrino mass leads to a falloff of the $e_1$ distribution 
at $K_{end} -m_\nu$ and a rise of the $e_2$ distribution at 
$K_{end} +m_\nu$, too fine a structure to see in the figure.

  We contrast the energy spectrum of $e_1$ from the
MDBD with that of the electron from the TBD. While the TBD produces a rapidly
decreasing electron energy spectrum as $K_e$ approaches $K_{end}$, 
the energy spectrum of $e_1$ rises gradually with $K_e$.  
Interestingly, the helicity factor
in the MDBD compensates the phase-space factor, leading to a much flatter
energy distribution for $e_1$ in MDBD.

\subsection{Background}

    Although the signal for a macroscopic neutrinoless double beta decay is that the sum of the energies of the two electrons in the event is
precisely $2K_{end}$, it would seem, given the high rates of single
beta decay, that the challenge of separating the signal from the 
background is insurmountable.   The $2\times10^{25}$  tritons in 100 g of 
tritium, with a half-life of
12.3 years,  would produce $\sim 2.5 \times 10^{16}$ decays per second.
However, only a tiny fraction of the decays give electrons with energy
near the beta decay endpoint.  An analogous challenge to
separate the signal from the background is encountered in the 
proposed detection of relic neutrinos using the ITBD reaction \cite{weinberg,
ptolemy}, where
the signals are separated from the endpoint by twice the neutrino mass. 
Long et al.~\cite{long} show that an energy resolution better than 0.7 $m_\nu$
is sufficient to reach a signal to background ratio better than 1.

    The signal to noise ratios for MDBD would be much better than in relic
neutrino detection, and do not require superb energy resolution.
As shown in Fig.~\ref{edistr}, one of the two electrons from MDBD has an energy
greater the endpoint energy of TBD, and up to twice the endpoint
energy. By requiring that the more energetic electron from the MDBD
candidate event has an energy sufficiently higher than the endpoint energy, the
background can be rejected at only a small cost to the MDBD detection
efficiency.

\begin{figure}[h]
\vspace{-48pt}
\includegraphics*[width=1.3\linewidth]{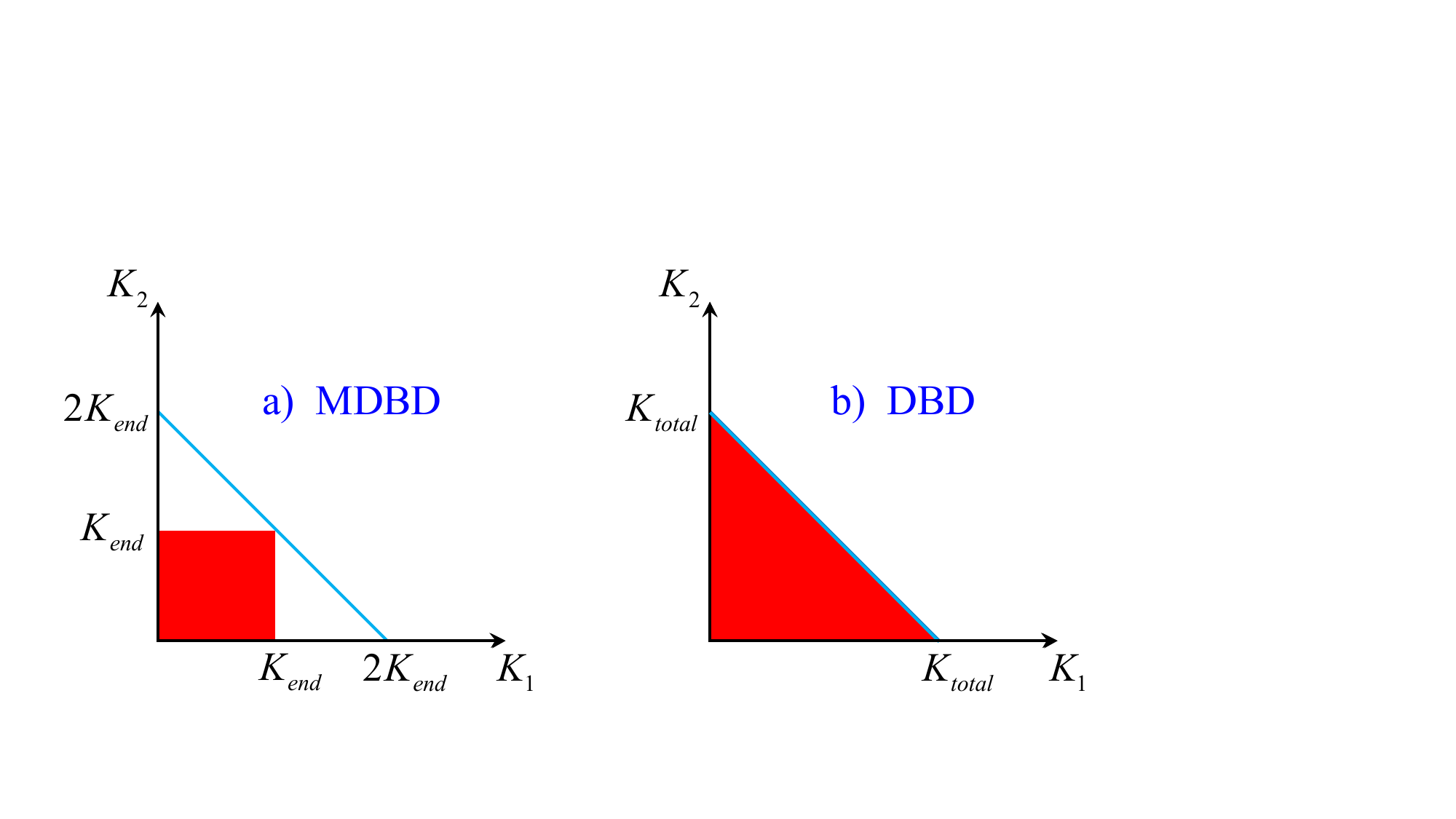}
\caption{The energy spectrum of two-electron events, with background in red, in a) MDBD and b) 0$\nu$DBD.   The axes are the kinetic energies of the individual electrons.  The kinetic energies of the two electrons in both neutrinoless double beta decays lie along the (blue) diagonal line extending from the first electron having $K_{total}$ (with  $K_{total}=2K_{end}$ in MDBD) and the second with 0,  to the reverse.  The energies of two electrons produced in single beta decays, the background in MDBD, lie in the red box in a), reaching the diagonal line only at a single point.  In contrast, in 0$\nu$DBD experiments the background from double beta decay with neutrinos lies in the red triangle in b), coming up to the diagonal line everywhere.}
\label{bkgnd}
\vspace{-24pt}
\end{figure}

   The favorable background rejection capability of MDBD is further
illustrated in Fig.~\ref{bkgnd}a, where the axes are the individual 
electron kinetic energies. An MDBD event would appear as a point
on the diagonal line $K_1+K_2 = \Delta M -2m_e$ in this figure.
The MDBD accidental background from two electrons emitted in two
independent single beta decays would 
lie in the red box in the figure, and except at
the single point where the red square touches the straight line, the background is
separated from the two electron MDBD event. The background from the
upper-right corner of the square in Fig.~\ref{bkgnd}a can be readily
rejected by requiring that the MDBD candidate events are well separated
from this corner.

    By contrast, the background in the $0\nu$DBD reaction  is
from the $2\nu$DBD reaction. 
The two electron kinetic energies from the $2\nu$DBD background
can reach the diagonal signal line in $0\nu$DBD everywhere, as shown 
by the red triangle in Fig.~\ref{bkgnd}b. These $2\nu$DBD background events
cannot be rejected without a significant loss of detection efficiency
for the $0\nu$DBD signals, unless superb energy resolution is achieved.

  The time separation between the two electrons can also be used to reject accidental background in MDBD.  The two 
electrons from an MDBD event  occur in a narrow time window, the travel time from source to target, $\delta t\sim R/c$, which can be of order tens of picoseconds, while the electrons from single beta decays are uncorrelated in time.   In contrast the $2\nu$DBD background in 0$\nu$DBD 
cannot be rejected on the basis of time separation.
 
As Kohyama et al.~\cite{Kohyama} pointed out,  solar 
neutrinos could be an important background for MDBD, since they could 
produce electrons with energies greater than the endpoint energy, just like the electrons from MDBD.     Indeed for single electron detection, the solar background is a serious concern.  However, solar neutrinos would have a totally negligible effect on the simultaneous detection
of two electrons produced in the MDBD, as considered in this paper, since
the probability for two electrons to be produced by the solar neutrinos at 
the same time or for a solar neutrino to produce an electron 
simultaneous with a single beta decay electron is extremely small.

\section{Conclusion}
\label{conclusion}

While neutrinoless double beta decay of single nuclei is the most promising experimental tool for testing the fundamental question of 
the Dirac or Majorana nature of neutrinos, the process of macroscopic neutrinoless double beta decay we have studied here is another manifestation of Majorana neutrinos with non-zero mass. 
The MDBD process shares features in common with single nucleus neutrinoless double beta decay; both depend on neutrinos being massive
and Majorana, and both are affected by quantum intererence between different neutrino mass states.   The two processes have significant differences, however, including the absence of nuclear matrix element uncertainties
for the MDBD rate and the different nature of the experimental backgrounds in these two processes.  In addition MDBD exhibits the new phenomenon of quantum coherence over macroscopic distances, related, but not equivalent to neutrino oscillations.

  Macroscopic double beta decay combines the processes of beta decay -- as is being studied for tritium in the KATRIN \cite{KATRIN-relic} experiment for measuring the 
neutrino mass -- and capture of neutrinos, as in the PTOLEMY experiment \cite{ptolemy} to detect primordial neutrinos from the Big Bang.    The MDBD process combining these two processes, could
in principle be accessible in either of these two experiments.   Our detailed analysis of various characteristics of MDBD, including expected rates, makes it clear that, given the quantities of tritium present, MDBD events are too rare to be detectable in the two experiments. 

   We emphasize again that MDBD in not a viable alternative, at this point, to the well established neutrinoless double beta decay experiments to test the Majorana nature of neutrinos.  We do not present MDBD as a currently feasible experiment.  Nevertheless, the similarities and distinct differences between these two processes provide useful perspectives on the underlying mechanisms for these two processes, and indicate new directions in which the concept of MDBD can be explored.   

    An example, as mentioned earlier, is the analog of MDBD that can occur with emission of a Majorana antineutrino with a flavor $\alpha$ other than electron, e.g., in 
 $\pi^-  \to \mu^- + \bar\nu_\mu$.   The amplitude for capture of the emitted antineutrino on a nucleus as an electron neutrino would be proportional to $\sum_i U_{\alpha i}U_{ei} m_{\nu i}$, which is non-zero even if no phases enter the $U_{\alpha i}$.   Such conversion of a Majorana $\bar\nu_\mu$ into a $\nu_e$ cannot occur in neutrinoless double beta decay within a single nucleus, simply by energy conservation, but can occur macroscopically.
This novel {\em flavor-changing} MDBD process,
a consequence of neutrino flavor mixing, is one example of how the MDBD
process discussed in this paper can be generalized into other physically
interesting processes.\footnote{We note that the similar process starting with  $\pi^+  \to \mu^+ + \nu_\mu$, in which the $\nu_\mu$ is later absorbed as a  $\nu_e$, is also flavor-changing, but does not require that neutrinos be Majorana.} 

Finally, while certain points discussed in this paper 
were first discussed in Refs.~\cite{Pacheco,Kohyama,Skalsey}, our 
present study  
contains a number of important new considerations not considered 
by these authors.  
We summarize these points:

\begin{itemize}

\item Section 3.1 discusses the physics of quantum interference
and coherence in the MDBD process. As we have shown, the rate of MDBD
depends on the neutrino mass hierarchy as well as the neutrino mixing
matrix.

\item While Refs.~\cite{Pacheco,Kohyama,Skalsey} only considered the 
overall rate of the MDBD, we have presented the expected energy spectra 
for the two electrons emitted in the MDBD process. We also show that the
electron energy spectrum for the electron produced in initial beta decay
for the MDBD process is very different from that in single
beta decay.

\item We have considered three specific cases, as shown in Table I,
for the expected yields for the MDBD process, and compared the
expected MDBD yields with the expected yields for the conventional
double-beta decays, using up-to-date nuclear matrix
elements. Such quantitative comparison between MDBD and DBD was not
attempted in Refs.~\cite{Pacheco,Kohyama,Skalsey}.

\item We have considered the MDBD process for a tritium target
and calculated the expected rate (Table 1), as well as the electron
energy spectrum. 
Since the ongoing KATRIN experiment and the proposed PTOLEMY experiment
are using tritium targets, our work provides needed input to estimate 
the expected rate of MDBD for these experiments.

\item We point out that an important experimental tool for separating
the MDBD signals from the single beta decay background is the 
simultaneous detection of both electrons emitted in the MDBD. 
References~\cite{Pacheco,Kohyama,Skalsey} only considered
the detection of of a single electron from MDBD. As we show in Fig. 6,
the detection of both electrons in the MDBD process can effectively reject the
accidental coincident background from single beta decay.
\end{itemize}

   We thank Prof.~Frank Deppisch and Dr. V. I. Tretyak for bringing the early work of A.~F. Pacheco to our attention.  This research was supported in part by the NSF Grant No. PHY-1812377 and by the Japan Science and Technology Agency (JST) as part of the Adopting Sustainable Partnerships for Innovative Research Ecosystem (ASPIRE), Grant Number JPMJAP2318, and was carried out in part at the Aspen Center for Physics, which is supported by National Science Foundation grant PHY-2210452, and at the National Central University in Taiwan under the Yushan Fellow Program.

\begin{appendix}
\label{gamma5}

\section{Helicity and chirality eigenstates eigenstates}
\label{gamma-hel} 

   Finite mass neutrinos can have positive in addition to negative helicity.    Here we derive the decomposition of helicity eigenstates into chiral eigenstates. 
   
    The Dirac state of a left handed (negative helicity) neutrino of energy $E$ and mass $m$, propagating in the +z direction say, is 
\beq
  u_{\nu L}  &=& \sqrt{\frac{E+m}{2E}}\left(0,1,0,-\frac{p}{E+m},0\right)^T \nonumber\\ &\equiv& (0,W_+,0,-W_-)^T, 
 \eeq
where $T$ denotes the transpose, and $W_\pm \equiv  \sqrt{(E\pm m)/2E}$.  From $W_+^2+W_-^2 = 1$ and $W_+W_-  = \beta_\nu/2$, we have $(W_+\pm W_-)^2 = 1 \pm \beta_\nu$, and 
\beq
 &&W_\pm  = \frac12\left( \sqrt{1+\beta}\pm\sqrt{1-\beta}\right). 
  \eeq
Similarly, the state of a right handed (positive helicity) neutrino is 
 \beq
  u_{\nu R} &=& \sqrt{\frac{E+m}{2E}}\left(1,0,\frac{p}{E+m},0\right)^T 
    \equiv (W_+,0,W_-,0)^T.
 \nonumber\\
 \eeq
 
    On the other hand, the $\gamma_5$ eigenstates for spin up or down (denoted by arrows) along the $z$-direction are
\beq
u_{\pm\uparrow} = \frac{(1,0,\pm1,0)^T}{\sqrt2}, \quad u_{\pm\downarrow} = \frac{(0,1,0,\pm1)^T}{\sqrt2}.
\eeq  
The helicity eigenstates can thus be written as
\beq
  u_{\nu R}   &=& \sqrt{\frac{1+\beta_\nu}{2}}u_{+\uparrow} + \sqrt{\frac{1-\beta\nu}{2}} u_{-\uparrow},
  \label{h1gamma}
\eeq
and
\beq
  u_{\nu L}  
  &=& \sqrt{\frac{1-\beta_\nu}{2}}u_{+\downarrow} + \sqrt{\frac{1+\beta_\nu}{2}}u_{- \downarrow}.
   \label{h-1gamma}
\eeq
The amplitude for a $\gamma_5=\pm 1$ eigenstate to have right handed helicity is $\sqrt{(1\pm\beta_\nu)/2}$ and to have left handed helicity is $\sqrt{(1\mp\beta_\nu)/2}$.   The expectation value of the neutrino helicity $\langle h\rangle $ in a $\gamma_5=\pm 1$ state is simply $\pm\beta_\nu$.

\end{appendix}

\end{document}